\documentclass[runningheads]{llncs}
\usepackage[T1]{fontenc}
\usepackage{booktabs}
\usepackage{graphicx}
\usepackage[most]{tcolorbox}
\usepackage{xcolor}
\usepackage{hyperref}

\definecolor{nsgteal}{HTML}{1D9E75}
\definecolor{nsgtealdark}{HTML}{04342C}
\definecolor{nsgtealfill}{HTML}{E1F5EE}
\definecolor{nsgcardbg}{HTML}{F4F8F6}

\newtcolorbox{finding}[1]{
  enhanced, breakable,
  colback=nsgcardbg, colframe=nsgcardbg,
  boxrule=0pt, leftrule=2.5pt,
  borderline west={2.5pt}{0pt}{nsgteal},
  arc=2pt, left=10pt, right=10pt, top=8pt, bottom=8pt,
  before upper={%
    \colorbox{nsgtealfill}{\textcolor{nsgtealdark}{\footnotesize\bfseries #1}}\par\vspace{4pt}}
}
\usepackage{enumitem}
\usepackage{fontawesome5}
\usepackage{multirow}

\newcommand{\CBS}{CBS}
\begin{document}
\title{Crossing the Cyber Divide: Sim-to-Sim and Sim-to-Real Transfer for RL Agents}
\author{}
\institute{}
\titlerunning{Crossing the Cyber Divide}
%
%
%
\author{Sabrina Saika\inst{1} \and
Yinuo Du\inst{1} \and
Aritran Piplai\inst{1}}

\authorrunning{S. Saika et al.}

\institute{The University of Texas at El Paso, El Paso, TX, USA \\
\email{ssaika@miners.utep.edu, \{ydu,apiplai\}@utep.edu}}

\maketitle

\begin{abstract}
Cyber attack agents are typically trained and evaluated within a single simulator, making it unclear whether learned policies transfer beyond the environments in which they were developed. This limitation hinders both deployment and fair comparison, as cyber simulators differ substantially in their state representations, observation models, and action spaces. In this paper, we study policy transfer across cyber environments and argue that simulator-to-simulator and simulator-to-real transfer can be viewed as instances of the same underlying alignment problem. We propose a framework that separates state alignment from action translation, enabling a policy trained in one environment to operate in another without retraining. We evaluate transfer across four cyber platforms, CyberBattleSim, NetSecGame, CyberWheel, and NASim, including emulated deployments in NASim. Our experiments show that zero-shot transfer is feasible, fully preserving source-policy performance in closely aligned environments and achieving ~45.2\% win rates when transferring policies whose source performance is ~60.5\%. In emulated virtual-machine environments, transferred policies exhibit a Jensen--Shannon divergence of 0.085 from native policies, indicating strong behavioral similarity. Code and benchmarks are available at:
\url{https://anonymous.4open.science/r/RL-Transfer-between-env-4F47/}. 

\keywords{Autonomous Cyber Agents  \and Reinforcement Learning \and AI-Driven Cyber Security}
\end{abstract}

\section{Introduction}
Autonomous cyber agents have emerged as a promising approach to scalable offensive security: enabling red-teaming, penetration testing, and attack-path discovery at a speed and breadth that human operators cannot match. Yet building agents that are both capable and deployable remains an open problem.

Large language model (LLM)-based agents, trained on massive corpora, offer strong few-shot adaptation and can reason about novel attack scenarios without task-specific retraining. LLM agents may successfully execute several steps of an offensive workflow before abruptly refusing to continue due to an internal safety mechanism. For autonomous red-teaming and penetration-testing tasks, such unpredictable compliance behavior can break an entire attack chain, making LLM-based agents powerful but difficult to deploy reliably. Heuristic and reinforcement learning (RL)-based agents, on the other hand, are controllable but brittle. Their decisions follow directly from interaction data and inspectable policies, but they overfit to the environments they are trained in. A policy trained on a specific network topology learns to exploit that topology's specifications, e.g., its subnet layout, service configuration, and vulnerability placement, and degrades sharply once they change. 
A second, \textit{less-resolved} limitation is that these agents act over abstract state and action spaces fixed at training time---host compromise flags, service enumeration counts, lateral-movement primitives---that have no counterpart in the concrete observations and tool invocations of a real deployment~\cite{kunz2023unified}. The simulators themselves compound the problem, since they differ substantially from one another, and a policy learned in one does not carry over to the next. This \textit{brittleness} hinders practitioners from (i) comparison and (ii) deployment. Comparing agents across simulators is problematic because performance often reflects simulator-specific assumptions rather than underlying policy quality. Deployment faces a related challenge: there is no agreed quantitative notion of fidelity for cyber environments~\cite{phase2025}, and building a high-fidelity replica for each target network is prohibitively expensive~\cite{kavak2021simulation}.

Consider an organization that wants to perform autonomous red-teaming safely. Because training directly on production infrastructure is unsafe, the agent is first trained in simulation. The organization then faces a problem: the simulator describes the world using its own observations and actions, which reflect the simulator's design rather than the organization's network. As a result, the trained policy cannot simply be deployed as-is. A similar problem arises when the organization wants to compare its agent against a state-of-the-art alternative trained on a different simulator. Even if both agents are solving the same task, they speak different representations of states and actions. In both cases, the challenge is how to move a learned policy from one environment to another without retraining it from scratch.

We view sim-to-sim and sim-to-real transfer as instances of the same underlying problem: aligning policies across environments with different state and action representations. The key distinction is that, in sim-to-real settings, the target environment is only partially known. If the full state, actions, and dynamics of the target were available, the problem would reduce to transferring between two fully specified environments. This does not make transfer easy; even when both environments are known, identifying an abstraction that relates them remains computationally difficult \cite{ravindran2003smdp}.
\textbf{This raises three questions: (RQ1)} Can offensive cyber policies transfer zero-shot across different environments without retraining? \textbf{(RQ2)} What aspects of a policy (e.g., states, actions) are actually transferable across environments? \textbf{(RQ3)} Does transfer remain possible when the target environment is only partially observable, as in emulated or real deployments? We evaluate these questions using a representation-alignment approach that separates state understanding from environment-specific action execution, across four environments: (i) \textit{CyberBattleSim}~\cite{cyberbattlesim}, (ii) \textit{NetSecGame}~\cite{netsecgame2023}, (iii) \textit{CyberWheel}~\cite{cyberwheel}, and (iv) \textit{NASim}~\cite{nasim}, providing both simulation and emulation.
\section{Related Work}
\subsection{Cyber Reinforcement Learning Agents}
Reinforcement learning-based agents have been one prominent type of autonomous cyber agent, alongside heuristic and LLM agents~\cite{redcase2023,pentestsurvey2026}. To train and evaluate these agents, many simulators have been developed~\cite{cyberbattlesim,standen2021cyborg,cyberwheel}. However, despite best efforts to conform to similar abstractions, the simulators are inevitably tightly coupled with their backend design. The differences in tech stack produce distributional differences in even the features they hold in common, such as observation dimensionality and action granularity. As a result, the field is siloed per simulator. Results do not compare across environments, and an agent or interface built for one does not carry to another.

\subsection{Policy Transfer Across Environments}
Policy transfer across environments asks whether a policy learned in one environment can be reused in another instead of trained from scratch. It is a natural solution to the fragmented landscape of cyber simulators. We face one of the harder cases in policy transfer with mismatch action and observation space, which places us in the cross-domain transfer regime ~\cite{you2022cross,serrano2023simknot}. Domain-adversarial alignment (DAPN)~\cite{ganin2016domain} is one of the most promising approach in this setting. It learns a feature space that preserves semantic classes while aligning across domains~\cite{gupta2017learning}. This maps naturally onto cyber offense, whose actions follow a shared kill-chain progression, which is captured by most simulators. We exploit this insight and operationalize DAPN for transfer across cyber simulators.

\subsection{Simulation-to-Real Transfer}
Simulation-to-Real (sim2real) transfer is a general challenge across robotics, autonomous driving, and control~\cite{zhao2020sim2real,simtorealsurvey2025} due to the cost and risk of real-world data~\cite{tobin2017domain}. In cybersecurity the challenges are even sharper for reasons such as the lack of faithful model of the network and adversary and user behavior. What's more, attempts to bridge the sim2real gap are expensive since live evaluation is costly and unsafe to run~\cite{pentestsurvey2026}. To bypass this constraint, we leverage the controllable sim2sim setting to gain insights about sim2real.

\subsection{Representation Alignment}
Representation alignment learns a shared feature space where inputs from different domains become comparable~\cite{ganin2015dann,transferability2022survey}. It has been applied across vision, NLP, and cross-domain control~\cite{ganin2016domain,you2022cross}. We cast cross-simulator cyber transfer as a representation alignment problem: the distinct observation spaces of two simulators become the domains to be aligned, into a shared representation a single policy can act on. This framing allows us to draw on a mature set of alignment methods and keeps our approach generally applicable.

\section{Methodology}

\begin{figure*}[t]
\centering
\includegraphics[width=\textwidth]{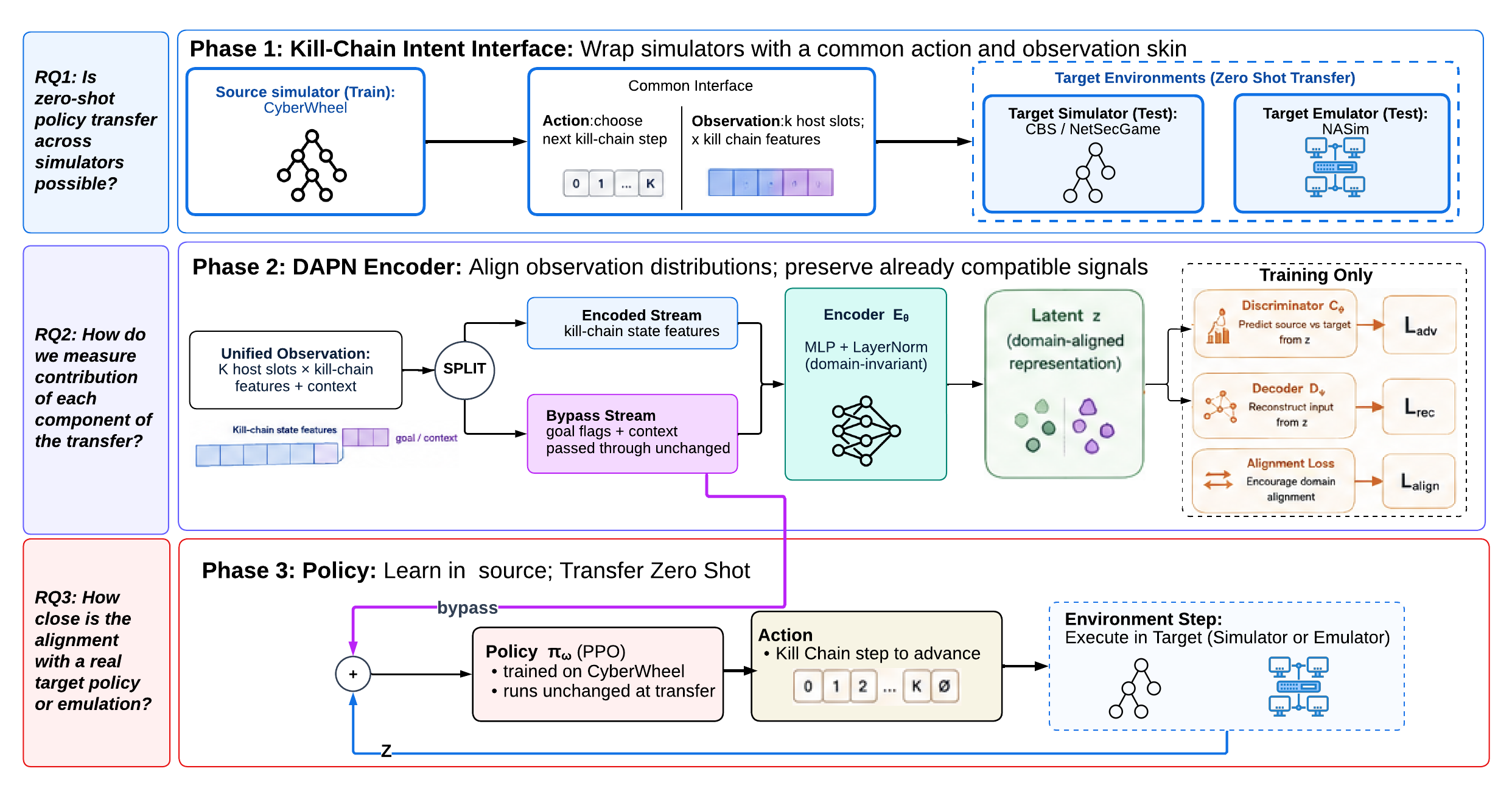}
\caption{Main architecture of the proposed framework. A policy trained in a source cyber environment is transferred to a target environment through state alignment and action translation, enabling zero-shot execution across simulators and emulated systems.}
\label{fig:architecture}
\end{figure*}

\subsection{Problem Statement}
 
Let $\mathcal{M}_A = \langle \mathcal{S}_A, \mathcal{A}_A, \mathcal{R}_A \rangle$ denote the environment defined by simulator $A$, where $\mathcal{S}_A$ is the state space, $\mathcal{A}_A$ is the action space, and $\mathcal{R}_A$ is the reward function. Let $\pi_A : \mathcal{S}_A \rightarrow \mathcal{A}_A$ be a policy trained to competence on $\mathcal{M}_A$. Given a target simulator $B$ with environment $\mathcal{M}_B = \langle \mathcal{S}_B, \mathcal{A}_B, \mathcal{R}_B \rangle$, we ask:\textit{ can $\pi_A$ be transferred to $\mathcal{M}_B$ without retraining? }
 
The core difficulty is that $\mathcal{S}_A \neq \mathcal{S}_B$ and $\mathcal{A}_A \neq \mathcal{A}_B$ in general, simulators differ in how they represent host state, network topology, and available actions, even when modeling semantically equivalent scenarios. A policy trained on $\mathcal{M}_A$ cannot be directly executed on $\mathcal{M}_B$ without a bridge between the two representations. We decompose the transfer problem into two sub-problems studied independently: 
(1) \textbf{State transfer:} Learn or construct a mapping $\phi: \mathcal{S}_B \rightarrow \mathcal{S}_A$ such that $\pi_A(\phi(s_B))$ produces meaningful actions when the agent observes $s_B \in \mathcal{S}_B$.
and 
(2) \textbf{Action transfer:} Learn or construct a mapping $\psi: \mathcal{A}_A \rightarrow \mathcal{A}_B$ such that the action selected by $\pi_A$ can be executed in $\mathcal{M}_B$. Figure~\ref{fig:architecture} shows our three-stage framework: (1) learn the action mapping $\psi$ via kill-chain transfer, (2) learn the state mapping $\phi$ using a modified DAPN adapter, and (3) use both mappings to deploy the source policy in the target environment.

\subsection{Kill-Chain Intent Interface}

We address the mismatch between $\mathcal{S}_A \neq \mathcal{S}_B$ and $\mathcal{A}_A \neq \mathcal{A}_B$ through a \emph{kill-chain intent interface}, a shared abstraction layer that constructs $\phi$ and $\psi$ without modifying the underlying simulators. In  Figure~\ref{fig:architecture} Phase 1, this interface is the central block between the two simulators, where $\phi$ (left) maps each observation space into the shared kill-chain representation and $\psi$ (right) maps
shared intents back into each simulator's action space.

\noindent\textbf{Action abstraction ($\psi$):}
The interface exposes a shared discrete action space in which each action selects a host to advance along the kill-chain, plus a no-op. The mapping $\psi$  resolves each selection into the appropriate simulator-native action, decoupling the policy's intent from the mechanical details of any given simulator.

\noindent\textbf{Observation abstraction ($\phi$):}
Each raw observation is projected into a fixed-size vector organized by kill-chain stage per tracked host, capturing phase, reachability, attacker presence, and target designation. A small context block appends episode-level signals such as the previous action and reward. Both $\phi$ and $\psi$ are computed deterministically via environment-specific wrappers; no learned components are involved.

\subsection{DAPN Encoder}
The unified interface constructs $\phi$ and $\psi$, resolving structural
mismatches between $\mathcal{S}_A$, $\mathcal{S}_B$ and $\mathcal{A}_A$,
$\mathcal{A}_B$. However, a distributional gap remains: even after applying
$\phi$, numerical values in the mapped state differ across simulators due to
different network sizes, discovery rates, and reward scales. 

We close this
gap with a modified \emph{DAPN}
encoder~\cite{zhao2020dapn} (Phase~2 in Figure~\ref{fig:architecture}),, which aligns the latent representations of
$\phi(\mathcal{S}_B)$ and  $\phi(\mathcal{S}_A)$ so that $\pi_A$ generalizes across
the domain boundary without retraining. Unlike the original DAPN architecture, an image-based few-shot classifier, our version uses a lightweight MLP encoder trained with adversarial domain confusion and reconstruction over cyber kill-chain features.

\noindent\textbf{Input partitioning:}
We split the observation into two streams before encoding. Features that vary across simulators (kill-chain state signals) pass through the encoder for alignment. Features that are semantically identical across domains (goal-identification flags) bypass the encoder entirely and are concatenated back at the output. Routing goal flags through adversarial alignment suppresses the goal signal; bypassing them preserves goal-directed behavior across the
domain boundary.

\noindent\textbf{Architecture:}
The encoder $E_\theta$ maps the variable-distribution stream to a compact latent vector through a shallow network with layer normalization. A paired decoder $D_\psi$ reconstructs the input from the latent, preventing the encoder from collapsing all observations to a single point under adversarial pressure. A binary discriminator $C_\phi$ predicts which simulator an observation came from; the encoder is trained to fool it.

\noindent\textbf{Training objective:}
The encoder is trained with three losses:
\begin{equation}
    \mathcal{L} \;=\;
    \lambda_{\mathrm{adv}}\,\mathcal{L}_{\mathrm{adv}}
    \;+\;
    \lambda_{\mathrm{rec}}\,\mathcal{L}_{\mathrm{rec}}
    \;+\;
    \lambda_{\mathrm{align}}\,\mathcal{L}_{\mathrm{align}}
    \label{eq:loss}
\end{equation}
$\mathcal{L}_{\mathrm{adv}}$ confuses the domain discriminator symmetrically
in both directions. $\mathcal{L}_{\mathrm{rec}}$ reconstructs observations
from both domains to prevent latent collapse. $\mathcal{L}_{\mathrm{align}}$
minimizes the distance between source and target latent means, stabilizing
early training. The encoder checkpoint is selected at chance discriminator
accuracy. Pretraining requires only random-policy observations from each
simulator, with no policy trajectories or target-domain rewards.

\section{Environments}
\label{sec:envs}

\begin{table}[t]
\centering
\caption{Environments and their characteristics}
\label{tab:envs}
\footnotesize
\setlength{\tabcolsep}{3pt}
\renewcommand{\arraystretch}{1.3}
\resizebox{\textwidth}{!}{%
\begin{tabular}{@{}l p{2.2cm} c p{2.4cm} p{2.6cm} p{2.4cm} p{2.8cm} p{2.2cm}@{}}
\toprule
& & \multicolumn{2}{c}{\textbf{Observation}} & \multicolumn{2}{c}{\textbf{Action}} & & \\
\cmidrule(lr){3-4}\cmidrule(lr){5-6}
\textbf{Env.} & \textbf{Train-to-converge} & \textbf{Dim} & \textbf{Decision-relevance} & \textbf{Space} & \textbf{Type} & \textbf{Task horizon} & \textbf{Configurable} \\
\midrule
CW & PPO steps (fast) & 78/512 & High (compact kill-chain) & 13 discrete & Informative + exploitative & 7 kill-chain phases/host & Yes (YAML hosts/subnets) \\
NSG & Fast ($\approx$ CW) & 78 & High (all features relevant) & 12 slot actions & Informative + exploitative & Discovery$\to$impact, $\leq$25 steps & Fixed topology \\
CBS & Slow & 512 & Low (many irrelevant features) & 9 discrete & Exploitative (discovery supplied) & Short: ${\sim}3$ actions, ${\sim}8$ steps/8 nodes & Yes, harder than CW \\
Emulation (NASim) & n/a & --- & n/a & Concrete tool invocations & Informative + exploitative & n/a & VMs, complex setup \\
\bottomrule
\end{tabular}%
}
\end{table}

We evaluate zero-shot policy transfer from a single source to two simulated
targets, with a dockerized emulation environment as a real-world proxy.
Table~\ref{tab:envs} summarizes observation and action structure, training
cost, task horizon, and configurability across environments.

\begin{itemize}[leftmargin=*, topsep=0pt, partopsep=0pt, parsep=0pt, itemsep=0pt]
    \item \noindent\textbf{CyberWheel} is the source domain. It is training-efficient,
    has a compact kill-chain observation with a high proportion of
    decision-relevant features, and is fully configurable via YAML.
    \item \noindent\textbf{CyberBattleSim} presents the larger domain gap. Its
    observation is high-dimensional and weakly structured; its action space
    is only partially aligned with CyberWheel. It is sample-inefficient to
    train but has a short task horizon once a policy exists.
    \item \noindent\textbf{NetSecGame} presents the smaller domain gap, with action
    and observation spaces closer to CyberWheel. Its task is harder to solve,
    requiring discovery followed by impact over a wider action set, lengthening
    the task horizon.
    \item \noindent\textbf{NASim} operates in simulation but also supports an
    emulation mode in which network nodes are instantiated as Docker containers
    or VMs, vulnerabilities are replicated on live services, and exploits are
    executed via Metasploit. This makes it a practical real-world proxy; full
    emulation results are deferred to future work.
\end{itemize}
 

\section{Experimental Setup}
 
 

\label{sec:experiments}
We evaluate zero-shot policy transfer from $\mathcal{M}_A$ to $\mathcal{M}_B$ using PPO. The source policy $\pi_A$ is trained in $\mathcal{M}_A$ and applied directly to $\mathcal{M}_B$ without retraining under four conditions: (1) \textbf{random policy}, a lower bound; (2) \textbf{zero-shot transfer} via index-aligned padding of states and actions, with no shared feature schema and no encoder, which tests whether a policy transfers under structurally misaligned observations; and (3) \textbf{feature-engineering transfer} via $\phi$ and $\psi$ (Table~\ref{tab:killchain}), which tests whether feature alignment alone suffices for zero-shot transfer.(4) \textbf{DAPN transfer}, which additionally applies $E_\theta$ to close the distributional gap between $\phi(\mathcal{S}_B)$ and $\mathcal{S}_A$. 
Configurations per
transfer pair are reported in Table \ref{tab:envs}.

\begin{table*}[t]
\centering
\small
\caption{Kill-chain phase mapping used by $\psi$ for the
CW$\rightarrow$\CBS, CW$\rightarrow$NSG, and CW$\rightarrow$NASim transfer pairs.}
\label{tab:killchain}
\resizebox{\textwidth}{!}{%
\begin{tabular}{clp{3.2cm}p{3.2cm}p{3.2cm}p{2.5cm}}
\toprule
\textbf{Ph.} & \textbf{Label} & \textbf{CBS $\psi$} & \textbf{NSG $\psi$} & \textbf{NASim $\psi$} & \textbf{Notes} \\
\midrule
0  & Network disc.  & \texttt{local\_exploit} \newline (frontier node) & \texttt{ScanNetwork()}             & \texttt{ServiceScan} \newline (subnet)        & Slot has no effect; $\psi$ is constant \\
1  & Service disc.  & \texttt{local\_exploit} \newline (frontier node) & \texttt{FindServices($h$)}         & \texttt{OSScan} / \newline \texttt{ProcScan}  & Slot has no effect; $\psi$ is constant \\
2  & Exploitation   & Connect \newline to target node         & \texttt{ExploitService($h$)}       & \texttt{SubnetScan}                           & Held until host is controlled \\
3 & Post-exploit & \texttt{local\_exploit} \newline (owned node)    & \texttt{FindData($h$)}             & \texttt{Exploit($h$)}                         & \\
4 & Exfiltration   & Cycles all vulnerabilities                       & \texttt{ExfiltrateData($h$, C\&C)} & \texttt{PrivEsc($h$) \newline \texttt{Exfiltrate Data}}                         & Requires data present \\
5+ & Terminal       & ---                                              & No-op                              & No-op                                         & NSG/NASim only \\
\bottomrule
\end{tabular}%
}
\end{table*}

\section{Results}
We present our findings by research question. We first validated that the DAPN encoder aligns the source and target distributions. Transfer is achievable but conditional on state alignment (RQ1); the aligned state representation carries it (RQ2); transferred policies track the emulator more closely than the source does (RQ3).

\begin{figure}[ht]
    \centering
    \includegraphics[width=\linewidth]{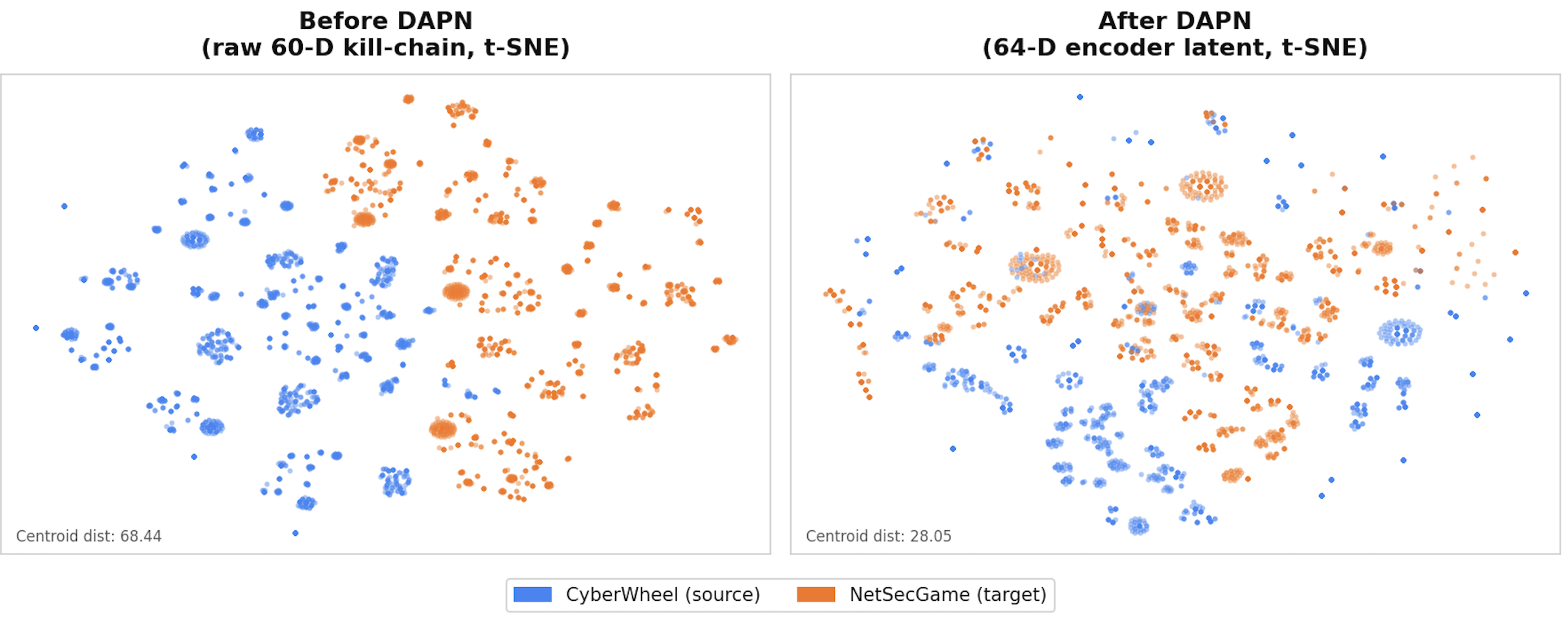}
    \caption{t-SNE projection of CW (blue) and NSG (orange) observations
             \emph{before} (left) and \emph{after} (right) the DAPN encoder.
             After encoding, the two domain clouds overlap substantially,
             confirming that the adversarial training closes the distributional
             gap between simulators.}
    \label{fig:latent-alignment}
\end{figure}
\begin{finding}{Latent alignment}
The DAPN encoder aligns source and target distributions at the latent level—centroid distance drops from \textbf{68.44 to 28.05}. 
\end{finding}
Figure~\ref{fig:latent-alignment} visualizes the effect of the DAPN encoder on
the observation distributions of the two simulators. Before encoding, the raw 60-D kill-chain features form two clearly separated clusters (centroid distance 68.44), reflecting the structural mismatch between CyberWheel and NetSecGame. After passing through the encoder, the 64-D latent representations of the two domains overlap substantially, with a centroid distance of 28.05 in the latent space. As the two distances are computed in different representation spaces, they are not directly comparable in magnitude; the substantial visual overlap in Figure~\ref{fig:latent-alignment} provides the primary evidence that the adversarial objective aligns the source and target distributions at the latent level.


\subsection{Can offensive cyber policies transfer zero-shot across different environments without retraining? (RQ1)}

\begin{finding}{NetSecGame · narrow gap target domain}
On a narrow domain gap, zero-shot transfer succeeds: DAPN reaches \textbf{45.2\%} target win rate, on par with feature engineering and far above the random (0.2\%) and unaligned (13.5\%) baselines.
\end{finding}
\vspace{-0.8cm}
\begin{table}[ht]
\centering
\caption{NSG evaluation results. Max 25 steps, 3 seeds $\times$ 500 episodes.}
\label{tab:nsg}
\resizebox{\textwidth}{!}{%
\begin{tabular}{lcccc}
\toprule
\textbf{Condition} & \textbf{CW Win\%} & \textbf{NSG Win\%}
                   & \textbf{Win Steps} & \textbf{Mean Return} \\
\midrule
Random Policy                          & ---     &  0.2\% & ---  & $-24.8$ \\
Zero-Shot Transfer      & 21.7\%  & 13.5\% & 7.3  & $-9.0$  \\
Zero-Shot Transfer + Feature Eng.      & 99.1\% & 47.4\% & 7.3  & $30.9$  \\
Zero-Shot Transfer + Feature Eng. + DAPN   & 60.5\%  & 45.2\% & 7.2  & $28.2$  \\
\bottomrule
\end{tabular}%
}
\end{table}
\vspace{-0.6cm}
Table~\ref{tab:nsg} presents the NSG evaluation results across 3 seeds (25-step budget, random target). The random policy wins only 0.2\% of episodes (1/500), confirming the task is non-trivial under unguided exploration, while unaligned zero-shot transfer reaches 13.5\% but stays at negative return ($-9.0$). Feature engineering lifts the win rate to 47.4\% ($+30.9$ return), and our domain-adversarial encoder preserves this target-domain performance (45.2\%, $+28.2$ return, 7.2 steps-to-win)---statistically on par.

\begin{finding}{CyberBattleSim · large gap target domain}
On a large domain gap, zero-shot transfer succeeds only with state alignment: DAPN reaches a \textbf{100\%} win rate. Both the random and feature-engineering baselines win \textbf{0\%}.
\end{finding}

\begin{table}[h]
  \centering
  \caption{Three conditions evaluated on CBS chain-12,
    win $=8$ nodes owned, $20$ episodes.}
  \label{tab:transfer-results}
  \resizebox{\columnwidth}{!}{%
  \begin{tabular}{lcccc}
    \toprule
    Condition   & Win Rate & Nodes Owned & Steps to 1st win & Mean Return \\
    \midrule
    Random Policy   & $0\%$            & $2.60$          & $21.9$          & $336$              \\
    Zero Shot + Feature Eng.      & $0\%$            & $0.00$          & ---             & $19$               \\
    \textbf{Zero Shot + Feature Eng. +DAPN} & $\mathbf{100\%}$ & $\mathbf{8.00}$ & $\mathbf{13.1}$ & $\mathbf{6{,}024}$ \\
    \bottomrule
  \end{tabular}%
  }
\end{table}
 
The results in Table~\ref{tab:transfer-results} are stark. The \textbf{Random Policy} baseline achieves a $0\%$ win rate despite unrestricted access to the full CBS action mask, owning on average only $2.60$ of the $8$ required nodes. This confirms that the chain-12 task is non-trivial under unguided exploration. \textbf{Zero-shot transfer with feature engineering alone} also wins $0\%$, with mean nodes owned collapsing to $0.00$ and mean return to $19$. \textbf{Zero-shot transfer with feature engineering and DAPN} achieves a $100\%$ win rate, reaching the $8$-node win condition in all $20$ episodes (mean return $6{,}024$; first compromise at $13.1$ steps on average). With $\phi$-encoded observations, the kill-chain translation layer and the learned state alignment together solve the task reliably.

\subsection{What aspects of a policy (e.g. states, actions) are actually transferable across environments? (RQ2)}

\begin{finding}{NetSecGame · narrow gap target domain}
The transferable aspect is the state representation: feature engineering overfits the source (99.1\% CW $\rightarrow$ 47.4\% NSG), while the encoder's representation holds across domains, reducing the source--target gap to \textbf{15.3 points} (70\% reduction).
\end{finding}

The NSG results in Table~\ref{tab:nsg} reveal which aspect of the policy carries the transfer. \textbf{Feature Engineering} nearly solves the source domain (99.1\% CW) yet loses over half its win rate on NSG, a 51.7-point drop that reveals heavy reliance on CW-specific structure. \textbf{DAPN} reduces the source-to-target gap to 15.3 points (60.5\% CW, 45.2\% NSG)---a 70\% reduction---yielding near domain-symmetric behaviour. Together with the collapse in cross-simulator distributional distance (Fig.~\ref{fig:latent-alignment}, centroid $68.44 \rightarrow 28.05$), this demonstrates that the adversarial encoder learns a genuinely domain-invariant representation rather than exploiting source-specific cues, achieving transfer that no longer presupposes the source and target are structurally identical.

\begin{finding}{CyberBattleSim · large gap target domain}
The transferable aspect is the distributionally-aligned state: matching observation shape alone leaves the policy emitting only no-ops (\textbf{0\%}), because the same feature index means different things in each environment.
\end{finding}

The CBS results in Table~\ref{tab:transfer-results} show that matching the action and feature layer is not the transferable aspect here---the state distribution is. \textbf{Zero-shot transfer with feature engineering alone} reshapes raw CBS observations into a $512$-dimensional vector, making them \emph{dimensionally} compatible with the CW-trained policy but not \emph{distributionally} compatible---the same feature index means something different in each environment. Faced with this mismatch, the policy's learned structure provides no usable signal: mean nodes owned collapses to $0.00$ and mean return to $19$, as the policy effectively emits only no-ops. Matching the shape of the observation space is therefore not enough; only aligning the state distribution bridges the two domains.

\begin{finding}{Transfer regimes}
On the small domain gap, feature engineering and DAPN reach comparable win rates (47.4\% and 45.2\%). On the large domain gap, feature engineering wins 0\% of episodes while DAPN wins 100\%.
\end{finding}

CBS and NSG span two ends of the transfer spectrum: a large source--target
schema gap (CBS) and a narrow one (NSG). Under the large gap (chain,
$500$ steps), alignment is decisive---DAPN wins $100\%$ while feature
engineering collapses to $0\%$, a degenerate failure in which the
raw-schema policy emits only no-ops. Under the narrow gap (random,
$25$ steps), the ordering shifts: Random ($0.2\%$) $<$ No Alignment
($13.5\%$) $<$ DAPN ($45.2\%$) $\approx$ Feature Eng. ($47.4\%$); here
feature engineering is competitive and the encoder adds no measurable gain,
even though DAPN's source policy is weaker ($60.5\%$ vs.\ $100\%$). The
results favor DAPN for robustness rather than peak performance: feature
engineering can match or exceed it when schemas align but fails outright
when they diverge, whereas DAPN produces useful policies in both regimes.

\subsection{Does transfer remain possible when the target environment is only partially observable, as in emulated or real deployments? (RQ3)}
\begin{finding}{Similarity with Emulator Policies}
Transferred policies behave more like the NASim emulator policy than the untransferred source does (\textbf{JS 0.087 / 0.085 vs 0.111}), supporting sim-to-real plausibility.
\end{finding}
\begin{figure}[ht]
  \centering
  \includegraphics[width=\textwidth]{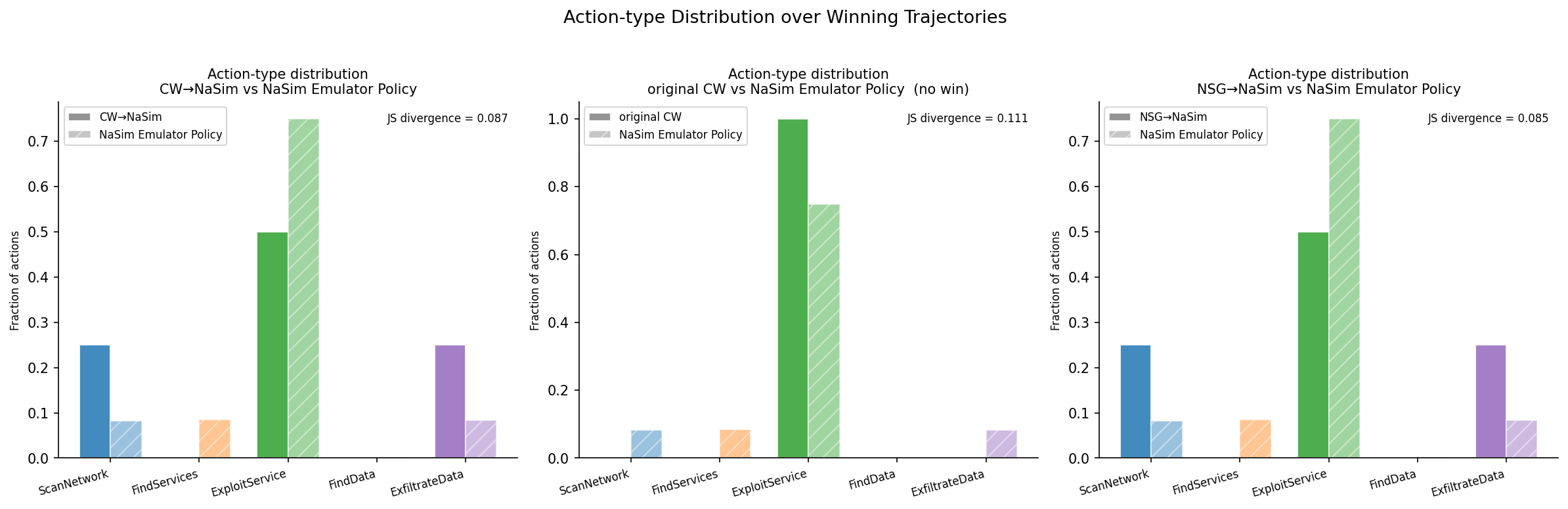}
  \caption{Action-type distribution over winning trajectories. Each panel
    compares a transferred policy against the NASim emulator policy across
    the five NASim action types with Jensen--Shannon divergence reported per
    panel. Translated CW$\rightarrow$NASim (JS $=0.087$). Original CW vs.
    NASim emulator policy, no win (JS $=0.111$). Translated NSG$\rightarrow$NASim
    (JS $=0.085$).}
  \label{fig:action-dist}
\end{figure}

Figure~\ref{fig:action-dist} shows the Jensen-Shannon divergence between
the action distributions of policies transferred to NASim and a reference
policy evaluated directly on the emulator. The emulator replicates the same
network topology as the simulator using Docker containers and VMs, with
vulnerabilities exploitable via Metasploit Framework.

\vspace{-1em}
\section{Conclusion and Future Work}

We present a method for zero-shot policy transfer across cybersecurity
simulators combining a kill-chain-based feature translator with a
domain-adaptive encoder. Together these components bridge both structural
and distributional gaps between environments without retraining the source
policy. Results show that when the source and target share a similar feature
schema, the translator alone is sufficient; the encoder becomes critical when
schemas diverge, recovering useful behavior where naive transfer collapses
entirely. Direct emulation remains an open challenge: beyond the
representational gap, real environments introduce practical obstacles such
as exploit version mismatches, indirect or delayed observation of action
outcomes, and non-deterministic service behavior that simulators abstract
away. Closing this sim-to-emulation gap is the natural next step, alongside
evaluation on branching attack topologies.

\vspace{-1em}
%
%
%
\bibliographystyle{splncs04}
\bibliography{ref}
\end{document}